\documentclass[12pt]{iopart}
\usepackage{graphicx}
\usepackage{dcolumn}
\usepackage{bm}
\usepackage{epsf}

\newcommand{\xiv}{\mbox{\boldmath$\xi$}}
\newcommand{\piv}{\mbox{\boldmath$\pi$}}
\newcommand{\alphav}{\mbox{\boldmath$\alpha$}}
\newcommand{\sigv}{\mbox{\boldmath$\sigma$}}

\begin{document}

\title{Coupling between a deuteron and a lattice}
\author{P L Hagelstein$^1$, I U Chaudhary$^2$}

\address{$^1$ Research Laboratory of Electronics, 
Massachusetts Institute of Technology, 
Cambridge, MA 02139,USA}
\ead{plh@mit.edu}

\address{$^2$ 
Department of Computer Science and Engineering, 
University of Engineering and Technology, 
Lahore, Pakistan}
\ead{irfanc@mit.edu}

\begin{abstract}
We recently put forth a new fundamental lattice Hamiltonian
based on an underlying picture of electrons and deuterons as elementary
Dirac particles.  Within this model there appears a term in which
lattice vibrations are coupled to internal nuclear transitions.  This
is interesting as it has the potential to provide a connection between
experiment and models that describe coherent energy transfer between two-level systems
and an oscillator.  In this work we describe a calculation of the coupling
matrix element in the case of the deuteron based on the old empirical 
Hamada-Johnston model for the nucleon-nucleon interaction.  The triplet S
and D states of the the deuteron in the rest frame couples to a singlet P
state through this new interaction.  The singlet P state in this calculation
is a virtual state with an energy of 125 MeV, and a coupling matrix element
for $z$-directed motion given by $2.98 \times 10^{-3} ~M_J c \hat{P}_z$.
\end{abstract}


\section{Introduction}

We recently obtained a new Hamiltonian for a lattice which includes interactions
with internal nuclear degrees of freedom \cite{Lattice}.  We started with a Dirac model for 
electrons and for nucleons on equal footing, then allowed the nuclei to
be described by a finite basis expansion, and finally developed an approximation
appropriate for low nuclear velocity.
Our original motivation for this was to obtain a model capable of describing the 
mass shift associated with excited nuclear states in a configuration interaction 
calculation.  
However, the new model unexpectedly contains a new coupling term which
provides for a direct interaction between lattice vibrations and nuclear transitions.

On the face of it, this new interaction term looks like it should allow for 
the strongly-coupled transitions that we have sought \cite{BirdsEye} in connection with the
generalized lossy spin-boson models \cite{CMNS5,CMNS7,CMNS9,CMNS10,CMNS11,Errata} that we proposed to account for some of
the anomalies (such as excess heat \cite{Flei89,Flei90,McKubre94,Storms} 
and collimated x-ray emission \cite{Karabut2002,Karabut2003,Karabut2004,Karabut2005,Karabut2012}) that have
been reported in experiments in recent years.  What is needed at this point
is an explicit calculation for some nuclear system to see how it works, what
states are coupled to, and how big the interaction is.

In general nuclear structure models are much more complicated than atomic structure
models due to the more complicated strong force interaction.  
We would like to work with empirical strong force models derived from scattering
experiments and few-body bound state binding energies.  In recent years these models
have achieved impressive results \cite{Kamada2001,Machleidt2011}; however, some of these strong force 
models involve a fair amount of work to implement.  
If we go back a few decades we
can find simpler versions of strong force models that are easier to work with, and
are sufficiently accurate to clarify the issues of interest here.  In the computations
that follow we will focus on the old Hamada-Johnston potential model \cite{Hamada1962}.
Without question the simplest compound nucleus which should show the effects of interest is
the deuteron, and so we will focus on this system in what follows.

In this formulation we have modeled the nucleons as elementary Dirac particles.  As nucleons
are made up of strongly interacting quarks, we know that they are not elementary Dirac particles.
To do better in the case of coupling with the deuteron, we would require a description 
in terms of the six constituent quarks.  We would expect from such a model a coupling
matrix element likely somewhat different from what we calculated in this work.
Even so, it makes sense here to pursue this simpler deuteron model based on 
simple Dirac nucleons as a step forward in the modeling process.


\newpage
\section{Basic model}

In a recent paper we discussed the derivation of a finite basis approximation
for a moving nucleus in the many-particle Dirac model which leads to the new
coupling that we are interested in.  We begin with the (relativistic)
finite basis model that we obtained.

\subsection{Finite basis approximation}

In \cite{Lattice} we developed finite basis eigenvalue
relations in the form

\begin{equation}
E c_n
~=~
\sqrt{ (M_nc^2)^2 + c^2 |{\bf P}|^2}
c_n
+
\sum_{m \ne n}
H_{nm} c_m
\end{equation}

\noindent
where the off-diagonal matrix elements were written as

\begin{equation}
H_{nm} 
~=~ 
\overline{\alphav}_{nm} \cdot 
 ( c  {\bf P} ) 
+
V_{nm}
\end{equation}

\noindent
Here $V_{nm}$ is the relative coupling matrix element

\begin{equation}
V_{nm}
~=~
\bigg \langle \Phi_n \bigg | 
\sum_j \alphav_j \cdot c \hat{\piv}_j 
+
\sum_j \beta_j m_jc^2
+
\sum_{j<k} \hat{V}_{jk}(\xiv_k-\xiv_j)
\bigg | \Phi_m \bigg \rangle
\end{equation}

\noindent
We defined $\overline{\alphav}_{nm}$ as

\begin{equation}
\overline{\alphav}_{nm}
~=~
\bigg \langle \Phi_n \bigg | 
 \sum_j {m_j \over M} \alphav_j  
\bigg | \Phi_m \bigg \rangle
\end{equation}


The notation for the two-body version of the problem is a bit different than
what we used for the many-particle problem.  It is useful to recast the
relative matrix element as

\begin{equation}
V_{nm}
~=~
\bigg \langle \Phi_n \bigg | 
(\alphav_2-\alphav_1) \cdot c \hat{\bf p} 
+
\beta_1 m_1c^2
+
\beta_2 m_2c^2
+
\hat{V}({\bf r})
\bigg | \Phi_m \bigg \rangle
\end{equation}

\noindent
The relative part of the off-diagonal matrix element corresponds to the rest frame
interaction terms, which might come about from strong force interactions as in
the development above for the nonrelativistic deuteron problem.  What is new is
the coupling with the center of mass momentum ${\bf P}$ that appears in 
$\overline{\alphav}_{nm} \cdot  ( c  {\bf P} )$.  We are interested in these
new matrix elements.

\subsection{Nonrelativistic reduction of the new interaction term}

In \cite{Lattice} we discussed the reduction of the new interaction
matrix element to the nonrelativistic case.  The results can be expressed as

\begin{samepage}

$$
\bigg \langle \Phi_f \bigg | 
 \sum_j {m_j \over M} \alphav_j  
\cdot c   \hat{\bf P} 
\bigg | \Phi_i \bigg \rangle
~\to~
{(E - M c^2) \over 2 Mc^2}
\bigg \langle \Phi_f \bigg | 
\sum_j { \hat{\piv_j} \cdot \hat{\bf P} \over m_j }
\bigg | \Phi_i \bigg \rangle
\ \ \ \ \ \ \ \ \ \ \ \ \ \ \ \ \ \ \ \ \ \ \ \ \ \ \ \ \
\ \ \ \ \ \ \ \ \ \ \ \ \ \ \ \ \ \ \ \ \ \ \ \ \ \ \ \ \
$$
$$
+
{1 \over 2 Mc^2}
\bigg [
\bigg \langle \Phi_f \bigg | 
\sum_j
(\sigv_j \cdot c   \hat{\bf P}) 
\bigg [
{1 \over 2 m_j c^2} 
\sum_{k<l} \hat{V}_{kl}(\xiv_l-\xiv_k)
\bigg ]
(\sigv_j \cdot c \hat{\piv}_j)
\bigg | \Phi_i \bigg \rangle
$$
\begin{equation}
+
\bigg \langle \Phi_f \bigg | 
\sum_j
(\sigv_j \cdot c   \hat{\piv}_j) 
\bigg [
{1 \over 2 m_j c^2} 
\sum_{k<l} \hat{V}_{kl}(\xiv_l-\xiv_k)
\bigg ]
(\sigv_j \cdot c \hat{\bf P})
\bigg | \Phi_i \bigg \rangle
\bigg ]
\end{equation}

\end{samepage}

\noindent
As above, this is written for the many-particle problem, and we wish to recast it
in terms of the two-body problem; we may write

$$
\bigg \langle \Phi_f \bigg | 
 \sum_j {m_j \over M} \alphav_j  
\cdot c   \hat{\bf P} 
\bigg | \Phi_i \bigg \rangle
~\to~
{(E - M c^2) \over 2 Mc^2}
\left ( {1 \over m_2} - {1 \over m_1} \right )
\bigg \langle \Phi_f \bigg | 
\hat{\bf p} \cdot \hat{\bf P}
\bigg | \Phi_i \bigg \rangle
\ \ \ \ \ \ \ \ \ \ \ \ \ \ \ \ \ \ \ \ \ \ \ \ \ \ \ \ \
\ \ \ \ \ \ \ \ \ \ \ \ \ \ \ \ \ \ \ \ \ \ \ \ \ \ \ \ \
$$
$$
-
{1 \over 2 Mc^2}
\bigg [
\bigg \langle \Phi_f \bigg | 
(\sigv_1 \cdot c   \hat{\bf P}) 
\bigg [
{1 \over 2 m_1 c^2} 
\hat{V}({\bf r})
\bigg ]
(\sigv_1 \cdot c \hat{\bf p})
\bigg | \Phi_i \bigg \rangle
$$
$$
+
\bigg \langle \Phi_f \bigg | 
(\sigv_1 \cdot c   \hat{\bf p}) 
\bigg [
{1 \over 2 m_1 c^2} 
\hat{V}({\bf r})
\bigg ]
(\sigv_1 \cdot c \hat{\bf P})
\bigg | \Phi_i \bigg \rangle
\bigg ]
$$
$$
+
{1 \over 2 Mc^2}
\bigg [
\bigg \langle \Phi_f \bigg | 
(\sigv_2 \cdot c   \hat{\bf P}) 
\bigg [
{1 \over 2 m_2 c^2} 
\hat{V}({\bf r})
\bigg ]
(\sigv_2 \cdot c \hat{\bf p})
\bigg | \Phi_i \bigg \rangle
$$
\begin{equation}
+
\bigg \langle \Phi_f \bigg | 
(\sigv_2 \cdot c   \hat{\bf p}) 
\bigg [
{1 \over 2 m_2 c^2} 
\hat{V}({\bf r})
\bigg ]
(\sigv_2 \cdot c \hat{\bf P})
\bigg | \Phi_i \bigg \rangle
\bigg ]
\end{equation}

\subsection{Equal mass approximation}

It is possible to split up this new interaction term into a contribution 
that takes the nucleon masses to be equal, and a small correction term
that depends on the difference between the nucleon masses.  In what follows
our focus will be on the larger equal mass terms, which is equivalent to 
making an equal mass approximation.  In this case we may write

\begin{samepage}

$$
\bigg \langle \Phi_f \bigg | 
 \sum_j {m_j \over M} \alphav_j  
\cdot c   \hat{\bf P} 
\bigg | \Phi_i \bigg \rangle
~\to~
-
\left ( {1 \over 2 Mc^2} \right )
\left ( {1 \over 2 m_{av}c^2} \right )
\bigg [
\bigg \langle \Phi_f \bigg | 
(\sigv_1 \cdot c   \hat{\bf P}) 
\bigg [
\hat{V}({\bf r})
\bigg ]
(\sigv_1 \cdot c \hat{\bf p})
\bigg | \Phi_i \bigg \rangle
\ \ \ \ \ \ \ \ \ \ \ \ \ \ \ \ \ \ \ \ \ \ \ \ \ \ \ \ \
\ \ \ \ \ \ \ \ \ \ \ \ \ \ \ \ \ \ \ \ \ \ \ \ \ \ \ \ \
$$
$$
+
\bigg \langle \Phi_f \bigg | 
(\sigv_1 \cdot c   \hat{\bf p}) 
\bigg [
\hat{V}({\bf r})
\bigg ]
(\sigv_1 \cdot c \hat{\bf P})
\bigg | \Phi_i \bigg \rangle
\bigg ]
$$
$$
+
\left ( {1 \over 2 Mc^2} \right )
\left ( {1 \over 2 m_{av}c^2} \right )
\bigg [
\bigg \langle \Phi_f \bigg | 
(\sigv_2 \cdot c   \hat{\bf P}) 
\bigg [
\hat{V}({\bf r})
\bigg ]
(\sigv_2 \cdot c \hat{\bf p})
\bigg | \Phi_i \bigg \rangle
$$
\begin{equation}
+
\bigg \langle \Phi_f \bigg | 
(\sigv_2 \cdot c   \hat{\bf p}) 
\bigg [
\hat{V}({\bf r})
\bigg ]
(\sigv_2 \cdot c \hat{\bf P})
\bigg | \Phi_i \bigg \rangle
\bigg ]
\end{equation}

\end{samepage}

\noindent
where we have assumed that

\begin{equation}
m_1 ~=~ m_2 ~=~ m_{av}
\end{equation}

\noindent
In this approximation there is no longer an explicit dependence on the state
energy $E$.

\subsection{Nonrelativistic approximation}

It is possible to develop a nonrelativistic approximation using

\begin{equation}
\sqrt{ (M_nc^2)^2 + c^2 |{\bf P}|^2}
~\to~
M_n c^2
+
{|{\bf P}|^2 \over 2 M_n}
\end{equation}

\noindent
In this case, a finite basis model that includes center of mass dynamics to lowest order
could be developed starting from a nonrelativistic Hamiltonian of the form

\begin{equation}
\hat{H}
~=~
\hat{M} c^2
+
{|\hat{\bf P}|^2 \over 2 \hat{M}}
+
{|\hat{\bf p}|^2 \over 2 \mu}
+
\hat{V}({\bf r})
+
\hat{\bf a} \cdot c \hat{\bf P}
\end{equation}

\noindent
where $\hat{M}$ is an operator that returns the rest mass energy
of the nuclear state, and
where $\hat{\bf a} \cdot c\hat{\bf P}$ in the equal mass approximation is

$$
\hat{\bf a} \cdot c\hat{\bf P}
~\to~
\left ( {1 \over 2 Mc^2} \right )
\left ( {1 \over 2 m_{av} c^2} \right ) 
\bigg [
(\sigv_2 \cdot c   \hat{\bf P}) 
\hat{V}
(\sigv_2 \cdot c \hat{\bf p})
+
(\sigv_2 \cdot c   \hat{\bf p}) 
\hat{V}
(\sigv_2 \cdot c \hat{\bf P})
\ \ \ \ \ \ \ \ \ \ \ \ \ \ \ \ \ \ \ \ \ \ \ \ \ \ \ \ \
\ \ \ \ \ \ \ \ \ \ \ \ \ \ \ \ \ \ \ \ \ \ \ \ \ \ \ \ \
$$
\begin{equation}
-
(\sigv_1 \cdot c   \hat{\bf P}) 
\hat{V}
(\sigv_1 \cdot c \hat{\bf p})
-
(\sigv_1 \cdot c   \hat{\bf p}) 
\hat{V}
(\sigv_1 \cdot c \hat{\bf P})
\bigg ]
\end{equation}


\newpage
\section{Finite-basis model for the deuteron}

We know from the literature that the deuteron at rest can be modeled using a triplet S
and triplet D state, since the tensor interaction mixes the two.  Since the kinetic energy
and potential terms preserve $J$ and $M_J$, each of the triplet S states mixes with 
a triplet D state that has the same $J$ and $M_J$.

\subsection{Mixing with $^1$P states}

The new interaction term causes these states to mix with singlet P states.  In general, the
new term does not preserve $M_J$, so that we would require a finite basis approximation that
distinguishes the different sublevels.
However, it is possible to focus on a special case of the new interaction which does preserve
$M_J$.  This occurs if we restrict our attention to

\begin{equation}
\hat{\bf P} ~=~ \hat{\bf i}_z \hat{P}_z
\end{equation}

\noindent
We find in this case that mixing occurs for $|M_J| = 1$, but not for $M_J = 0$.  In response, we
might write

\begin{equation}
\Psi ~=~ \Psi_{^3{\rm S}} + \Psi_{^3{\rm D}} + \Psi_{^1{\rm P}}
\end{equation}

\noindent
with the understanding that 

\begin{equation}
\Psi_{^1{\rm P}}(J_M = 0) ~\to~ 0
\end{equation}

\subsection{Basis state construction}

Nuclear state construction is usually carried out in the isospin scheme, with antisymmetry enforced
through the application of the generalized Pauli principle.  The two-body problem is particularly
simple in this regard, with spin, isospin and spatial components restricted to being either symmetric $(s)$
or antisymmetric $(a)$; we may write for the three states

\begin{equation}
\Psi_{^3{\rm S}} ~=~ R(s) S(s) T(a)
\end{equation}

\begin{equation}
\Psi_{^3{\rm D}} ~=~ R(s) S(s) T(a)
\end{equation}

\begin{equation}
\Psi_{^1{\rm P}} ~=~ R(a) S(a) T(a)
\end{equation}

\noindent
The antisymmetric spin and isospin terms $S(a)$ and $T(a)$ are singlets, and the symmetric
spin and isospin terms $S(s)$ and $T(s)$ are triplets.

\subsection{Triplet S state}

The S state is a triplet spin state, so we may write
it as

$$
\Psi_{^3{\rm S}} ~=~
\Psi_{^3{\rm S}} (S=1,M_S;T=0,M_T=0;l=0,m=0) 
$$
\begin{equation}
~=~ 
{u(r) \over r} Y_{00}(\theta,\phi)~ |1,M_S \rangle_S~ |0,0\rangle_T
\end{equation}

\noindent
The $Y_{lm}(\theta,\phi)$ are spherical harmonics; we choose $l=0$ and $m=0$ since
we are working with an S state.  The $|S,M_S \rangle_S$ are spin functions for the neutron
and proton spins; the $|T,M_T\rangle_T$ are isospin functions, and we have used an
isospin singlet function here.

\subsection{Triplet D state}

We can develop a D state by applying the tensor $\hat{S}_{12}$ operator on an S state.
This approach was used early on as a convenient way of generating few-body wavefunctions
for variational calculations in nuclear physics.  We may write

\begin{equation}
\Psi_{^3{\rm D}} ~=~
{1 \over \sqrt{8}}
\hat{S}_{12} \left [{v(r) \over r} Y_{00}(\theta,\phi) |1,M_S \rangle_S~ |0,0\rangle_T \right ]
\end{equation}

\noindent
This construction is convenient since

\begin{equation}
\hat{S}_{12} \Psi_{^3{\rm D}}
~=~
\sqrt{8} \left [{v(r) \over r} Y_{00}(\theta,\phi) |1,M_S \rangle_S~ |0,0\rangle_T \right ]
-
2 \Psi_{^3D}
\end{equation}

\subsection{Singlet P state}

The singlet P state for a particular calculation can be specified using

$$
\Psi_{^1{\rm P}} ~=~
\Psi_{^1{\rm P}} (S=0,M_S=0;T=0,M_T=0;l=1,m) 
$$
\begin{equation}
~=~ 
i {w(r) \over r} Y_{1m}(\theta,\phi)~ |0,0 \rangle_S~ |0,0\rangle_T
\end{equation}

\noindent
Including an $i$ here leads to real coupling coefficients in what follows.

\subsection{Normalization}

We can evaluate the normalization integral for these states simply; we write

$$
\langle \Psi | \Psi \rangle
~=~
\langle \Psi_{^3{\rm S}} | \Psi_{^3{\rm S}} \rangle
+
\langle \Psi_{^3{\rm D}} | \Psi_{^3{\rm D}} \rangle
+
\langle \Psi_{^1{\rm P}} | \Psi_{^1{\rm P}} \rangle
$$
\begin{equation}
~=~
\int_0^\infty
|u(r)|^2
+
|v(r)|^2
+
|w(r)|^2
dr
\end{equation}

\subsection{Expectation value of Hamiltonian terms}

We are interested in developing coupled channel equations that include the new interaction.  For
the problem in the rest frame, this is most easily accomplished by developing an expression
for the total energy and then using the variational principle.  We can use the same basic 
approach here for the moving frame version of the problem.
We begin with

$$
\left \langle \Psi \left |{{\bf p}^2 \over 2 \mu} + \hat{V} + ({\bf a} \cdot c \hat{\bf P})_z \right | \Psi \right \rangle
~=~
\ \ \ \ \ \ \ \ \ \ \ \ \ \ \ \ \ \ \ \ \ \ \ \ \ \ \ 
\ \ \ \ \ \ \ \ \ \ \ \ \ \ \ \ \ \ \ \ \ \ \ \ \ \ \ 
\ \ \ \ \ \ \ \ \ \ \ \ \ \ \ \ \ \ \ \ \ \ \ \ \ \ \ 
\ \ \ \ \ \ \ \ \ \ \ \ \ \ \ \ \ \ \ \ \ \ \ \ \ \ \ 
$$
$$
\left \langle \Psi_{^3{\rm S}} \left |{{\bf p}^2 \over 2 \mu} + \hat{V} \right | \Psi_{^3{\rm S}} \right \rangle
+
\left \langle \Psi_{^3{\rm D}} \left |{{\bf p}^2 \over 2 \mu} + \hat{V} \right | \Psi_{^3{\rm D}} \right \rangle
+
\left \langle \Psi_{^1{\rm P}} \left |{{\bf p}^2 \over 2 \mu} + \hat{V} \right | \Psi_{^1{\rm P}} \right \rangle
$$
$$
+
\left \langle \Psi_{^3{\rm S}} \left | \hat{V} \right | \Psi_{^3{\rm D}} \right \rangle
+
\left \langle \Psi_{^3{\rm D}} \left | \hat{V} \right | \Psi_{^1{\rm S}} \right \rangle
+
\left \langle \Psi_{^1{\rm P}} \left | ({\bf a} \cdot c \hat{\bf P})_z \right | \Psi_{^1{\rm S}} \right \rangle
$$
\begin{equation}
+
\left \langle \Psi_{^1{\rm P}} \left | ({\bf a} \cdot c \hat{\bf P})_z \right | \Psi_{^3{\rm D}} \right \rangle
+
\left \langle \Psi_{^3{\rm S}} \left | ({\bf a} \cdot c \hat{\bf P})_z \right | \Psi_{^1{\rm P}} \right \rangle
+
\left \langle \Psi_{^3{\rm D}} \left | ({\bf a} \cdot c \hat{\bf P})_z \right | \Psi_{^1{\rm P}} \right \rangle
\end{equation}

\subsection{Diagonal matrix elements}

We can evaluate the diagonal matrix elements directly using Mathematica to obtain

\begin{equation}
\left \langle \Psi_{^3{\rm S}} \left |{{\bf p}^2 \over 2 \mu} + \hat{V} \right | \Psi_{^3{\rm S}} \right \rangle
~=~
\int_0^\infty u(r) \bigg [  -{\hbar^2 \over 2 \mu}{d^2 \over dr^2} - 3v_C^{et}(r) \bigg ] u(r) dr
\end{equation}

\begin{equation}
\left \langle \Psi_{^3{\rm D}} \left |{{\bf p}^2 \over 2 \mu} + \hat{V} \right | \Psi_{^3{\rm D}} \right \rangle
~=~
\int_0^\infty v(r) \bigg [ -{\hbar^2 \over 2 \mu}{d^2 \over dr^2} - 3v_C^{et}(r) + 6v_T^{et}(r) -3v_{LS}^{et}(r) - 3 v_{LL}^{et}(r) \bigg ] v(r)dr
\end{equation}

\begin{equation}
\left \langle \Psi_{^1{\rm P}} \left |{{\bf p}^2 \over 2 \mu} + \hat{V} \right | \Psi_{^1{\rm P}} \right \rangle
~=~
\int_0^\infty w(r) \bigg [ -{\hbar^2 \over 2 \mu}{d^2 \over dr^2} + 9v_C^{os}(r)   - 2 v_{LL}^{os}(r) \bigg ] w(r) dr
\end{equation}

\subsection{Off-diagonal potential matrix elements}

In the case of the Hamada-Johnston potential, there occur off-diagonal matrix elements between the triplet S and singlet D states,
which are given by

\begin{equation}
\left \langle \Psi_{^3{\rm S}} \left | \hat{V} \right | \Psi_{^3{\rm D}} \right \rangle 
~=~
\left \langle \Psi_{^3{\rm D}} \left | \hat{V} \right | \Psi_{^3{\rm S}} \right \rangle 
~=~
-3\sqrt{8} \int_0^\infty u(r)  v_{T}^{et} v(r) dr 
\end{equation}

\noindent
The superscript $et$ in the associated potentials here is connected with the even triplet channel, since the Hamada-Johnston
potentials are fit for the different channels separately.

\subsection{Off-diagonal matrix elements for the new interaction}

For the off-diagonal matrix elements of the new interaction, we have used Mathematica to compute

$$
\langle \Phi_{^1{\rm P}} | ({\bf a} \cdot c \hat{\bf P})_z | \Phi_{^3{\rm S}} \rangle
~=~
M_J
\left ( {1 \over 2 Mc^2} \right )
\left ( {1 \over 2 m_{av} c^2} \right )
(\hbar c) (c \hat{P}_z)
\ \ \ \ \ \ \ \ \ \ \ \ \ \ \ \ \ \ \ \ \ \ \ \ \ \ \ \ 
\ \ \ \ \ \ \ \ \ \ \ \ \ \ \ \ \ \ \ \ \ \ \ \ \ \ \ \ 
\ \ \ \ \ \ \ \ \ \ \ \ \ \ \ \ \ \ \ \ \ \ \ \ \ \ \ \ 
\ \ \ \ \ \ \ \ \ \ \ \ \ \ \ \ \ \ \ \ \ \ \ \ \ \ \ \ 
$$
$$
\bigg \lbrace
-2 \sqrt{3}   \int_0^\infty w(r) \left [{d \over dr} v_C^{eT}(r) \right ] u(r) dr
$$
$$
+
 12 \sqrt{3}  \int_0^\infty w(r) v_T^{eT}(r) \left [ {d \over dr} u(r) + {u(r) \over r} \right ] dr
$$
$$
+
 8 \sqrt{3}   \int_0^\infty w(r) \left [{d \over dr} v_T^{eT}(r)  \right ] u(r) dr
$$
$$
+
{2 \over \sqrt{3}}   \int_0^\infty w(r) v_{LS}^{eT}(r) \left [ {d \over dr} u(r) - {u(r) \over r} \right ] dr
$$
\begin{equation}
- 2 \sqrt{3}  \int_0^\infty w(r) v_{LL}^{eT}(r) \left [ {d \over dr} u(r) - {u(r) \over r} \right ] dr
\bigg \rbrace
\end{equation}

$$
\langle \Phi_{^1{\rm P}} | ({\bf a} \cdot c \hat{\bf P})_z | \Phi_{^3{\rm D}} \rangle
~=~
M_J
\left ( {1 \over 2 Mc^2} \right )
\left ( {1 \over 2 m_{av} c^2} \right )
(\hbar c) (c \hat{P}_z)
\ \ \ \ \ \ \ \ \ \ \ \ \ \ \ \ \ \ \ \ \ \ \ \ \ \ \ \ 
\ \ \ \ \ \ \ \ \ \ \ \ \ \ \ \ \ \ \ \ \ \ \ \ \ \ \ \ 
\ \ \ \ \ \ \ \ \ \ \ \ \ \ \ \ \ \ \ \ \ \ \ \ \ \ \ \ 
\ \ \ \ \ \ \ \ \ \ \ \ \ \ \ \ \ \ \ \ \ \ \ \ \ \ \ \ 
$$
$$
\bigg \lbrace
- \sqrt{6}   \int_0^\infty w(r) \left [{d \over dr} v_C^{eT}(r) \right ] v(r) dr
$$
$$
+ 6 \sqrt{6}  \int_0^\infty w(r) v_T^{eT}(r) \left [ {d \over dr} v(r) + {v(r) \over r} \right ] dr
$$
$$
+ 4 \sqrt{6}  \int_0^\infty w(r) \left [{d \over dr} v_T^{eT}(r) \right ] v(r) dr
$$
$$
- \sqrt{8 \over 3}   \int_0^\infty w(r) v_{LS}^{eT}(r) \left [ {d \over dr} v(r) + 2{v(r) \over r} \right ] dr
$$
$$
- \sqrt{6}  \int_0^\infty w(r) \left [ {d \over dr} v_{LS}^{eT}(r) \right ] v(r)  dr
$$
$$
- 2 \sqrt{6} \int_0^\infty w(r) v_{LL}^{eT}(r) \left [ {d \over dr} v(r) + 2{v(r) \over r} \right ] dr
$$
\begin{equation}
- \sqrt{6}  \int_0^\infty w(r) \left [ {d \over dr} v_{LL}^{eT}(r) \right ] v(r)  dr
\bigg \rbrace
\end{equation}

$$
\langle \Phi_{^3{\rm S}} | ({\bf a} \cdot c \hat{\bf P})_z | \Phi_{^1{\rm P}} \rangle
~=~
M_J
\left ( {1 \over 2 Mc^2} \right )
\left ( {1 \over 2 m_{av} c^2} \right )
(\hbar c) (c \hat{P}_z)
\ \ \ \ \ \ \ \ \ \ \ \ \ \ \ \ \ \ \ \ \ \ \ \ \ \ \ \ 
\ \ \ \ \ \ \ \ \ \ \ \ \ \ \ \ \ \ \ \ \ \ \ \ \ \ \ \ 
\ \ \ \ \ \ \ \ \ \ \ \ \ \ \ \ \ \ \ \ \ \ \ \ \ \ \ \ 
\ \ \ \ \ \ \ \ \ \ \ \ \ \ \ \ \ \ \ \ \ \ \ \ \ \ \ \ 
$$
$$
\bigg \lbrace
- 2 \sqrt{3}   \int_0^\infty u(r) \left [{d \over dr} v_C^{eT}(r) \right ] w(r) dr
$$
$$
+
 12 \sqrt{3}  \int_0^\infty u(r) v_T^{eT}(r) \left [ -{d \over dr} w(r) + {w(r) \over r} \right ] dr
$$
$$
- 
4 \sqrt{3}   \int_0^\infty u(r) \left [{d \over dr} v_T^{eT}(r)  \right ] w(r) dr
$$
$$
+
{2 \over \sqrt{3}}   \int_0^\infty w(r) v_{LS}^{eT}(r) \left [ -{d \over dr} u(r) - {u(r) \over r} \right ] dr
$$
$$
- 
{2 \over \sqrt{3}}   \int_0^\infty u(r) \left [{d \over dr} v_{LS}^{eT}(r)  \right ] w(r) dr
$$
$$
- 2 \sqrt{3}  \int_0^\infty w(r) v_{LL}^{eT}(r) \left [ -{d \over dr} u(r) - {u(r) \over r} \right ] dr
$$
\begin{equation}
+ 2 \sqrt{3}   \int_0^\infty u(r) \left [{d \over dr} v_{LL}^{eT}(r)  \right ] w(r) dr
\bigg \rbrace
\end{equation}

$$
\langle \Phi_{^3{\rm D}} | ({\bf a} \cdot c \hat{\bf P})_z | \Phi_{^1{\rm P}} \rangle
~=~
M_J
\left ( {1 \over 2 Mc^2} \right )
\left ( {1 \over 2 m_{av} c^2} \right )
(\hbar c) (c \hat{P}_z)
\ \ \ \ \ \ \ \ \ \ \ \ \ \ \ \ \ \ \ \ \ \ \ \ \ \ \ \ 
\ \ \ \ \ \ \ \ \ \ \ \ \ \ \ \ \ \ \ \ \ \ \ \ \ \ \ \ 
\ \ \ \ \ \ \ \ \ \ \ \ \ \ \ \ \ \ \ \ \ \ \ \ \ \ \ \ 
\ \ \ \ \ \ \ \ \ \ \ \ \ \ \ \ \ \ \ \ \ \ \ \ \ \ \ \ 
$$
$$
\bigg \lbrace
 - \sqrt{6}   \int_0^\infty v(r) \left [{d \over dr} v_C^{eT}(r) \right ] 2(r) dr
$$
$$
+ 6 \sqrt{6}  \int_0^\infty v(r) v_T^{eT}(r) \left [ -{d \over dr} w(r) + {w(r) \over r} \right ] dr
$$
$$
- 2 \sqrt{6}  \int_0^\infty v(r) \left [{d \over dr} v_T^{eT}(r) \right ] w(r) dr
$$
$$
- \sqrt{8 \over 3}   \int_0^\infty w(r) v_{LS}^{eT}(r) \left [ -{d \over dr} v(r) + 2{v(r) \over r} \right ] dr
$$
$$
- \sqrt{2 \over 3}  \int_0^\infty w(r) \left [ {d \over dr} v_{LS}^{eT}(r) \right ] v(r)  dr
$$
$$
- 2 \sqrt{6} \int_0^\infty w(r) v_{LL}^{eT}(r) \left [ -{d \over dr} v(r) + 2{v(r) \over r} \right ] dr
$$
\begin{equation}
+ \sqrt{6}  \int_0^\infty w(r) \left [ {d \over dr} v_{LL}^{eT}(r) \right ] v(r)  dr
\bigg \rbrace
\end{equation}


\newpage
\section{Coupled-channel equations}

We have specified a finite basis problem with three channels, which would produce
three complicated coupled-channeled equations if we decided to treat the different
basis states on equal footing.  However, since the momentum ${\bf P}$ that we 
are interested in for applications of this model is small, the triplet S and D
channels are then best considered to constitute the unperturbed deuteron problem,
and the singlet P channel will contain the weak response of the deuteron to
the ${\bf a} \cdot c {\bf P}$ perturbation.

In this case, it seems appropriate to develop the coupled triplet S and D channels
consistent with the rest frame deuteron problem.  Once the associated wavefunctions
are known, then we can use them to approximate the occupation of the singlet P channel.


\subsection{Rarita-Schwinger equations}

Given the approach outlined above, we can optimize the channel wavefunctions $u(r)$ and $v(r)$ by
minimizing the rest frame energy

$$
\left \langle \Psi_{^3{\rm S}} \left |{{\bf p}^2 \over 2 \mu} + \hat{V} \right | \Psi_{^3{\rm S}} \right \rangle
+
\left \langle \Psi_{^3{\rm D}} \left |{{\bf p}^2 \over 2 \mu} + \hat{V} \right | \Psi_{^3{\rm D}} \right \rangle
+
\left \langle \Psi_{^3{\rm S}} \left | \hat{V} \right | \Psi_{^3{\rm D}} \right \rangle 
+
\left \langle \Psi_{^3{\rm D}} \left | \hat{V} \right | \Psi_{^3{\rm S}} \right \rangle 
~=~
$$
$$
\int_0^\infty u(r) \bigg [  -{\hbar^2 \over 2 \mu}{d^2 \over dr^2} - 3v_C^{et}(r) \bigg ] u(r) dr
$$
$$
+
\int_0^\infty v(r) \bigg [ -{\hbar^2 \over 2 \mu}{d^2 \over dr^2} + {6\hbar^2 \over 2\mu r^2} - 3v_C^{et}(r) + 6v_T^{et}(r) -3v_{LS}^{et}(r) - 3 v_{LL}^{et}(r) \bigg ] v(r)dr
$$
\begin{equation}
-6\sqrt{8} \int_0^\infty u(r)  v_{T}^{et} v(r) dr 
\end{equation}

\noindent
The minimization of this rest frame energy leads to the constraints

\begin{equation}
E_r u(r) 
~=~
\left [ -{\hbar^2 \over 2 \mu}{d^2 \over dr^2} - 3v_C^{et}(r) \right ] u(r)
 + \bigg [ - 3\sqrt{8} v_{T}^{et}(r) \bigg ] v(r)
\end{equation}

$$
E_r v(r)
~=~
\left [
 -{\hbar^2 \over 2 \mu}{d^2 \over dr^2} +{6 \hbar^2 \over 2 \mu r^2} - 3v_C^{et}(r) + 6v_T^{et}(r) -3v_{LS}^{et}(r) - 3 v_{LL}^{et}(r)
\right ] v(r)
$$
\begin{equation}
 + \bigg [ - 3\sqrt{8} v_{T}^{et}(r) \bigg ] u(r)
\end{equation}

\noindent
where $E_r$ is the relative energy.
We recognize these as the Rarita-Schwinger equations based on the Hamada-Johnston potential model.

\subsection{Rest frame triplet S and D channel wavefunctions}

\begin{figure} [t]
\begin{center}
\mbox{\epsfbox{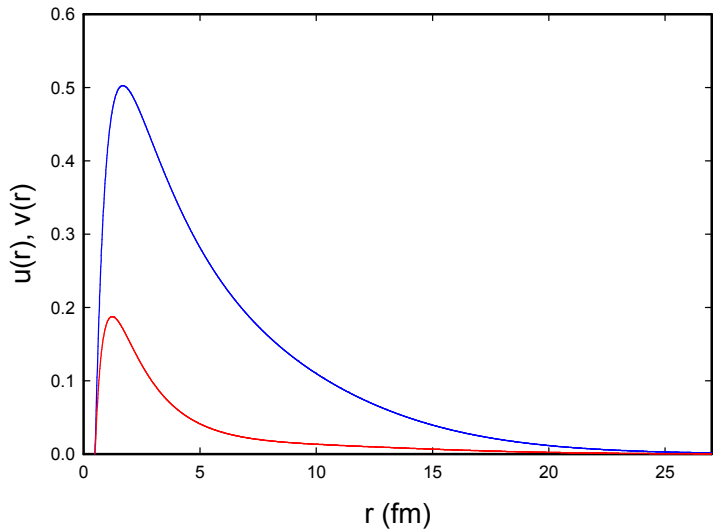}}
\caption{Numerical solutions of the Rarita-Schwinger equations for the deuteron
using the Hamada-Johnston potential. The solution for the S state [$u(r)$] is shown in blue; the solution
for the D state [$v(r)$] is shown in red.}
\label{deuteron1}
\end{center}
\end{figure}

We have solved the Rarita-Schwinger equations to obtain the channel wavefunctions plotted in Figure \ref{deuteron1}.
The triplet S channel wavefunction $u(r)$ is larger and extends out to a relatively large radial separation, and
the triplet D channel wavefunction $v(r)$ is smaller and localized to much smaller radial separation.  We can see
the effect of the hard core potential in the zero boundary condition at the cut off radius.


\epsfxsize = 4.00in
\epsfysize = 3.200in

\subsection{Optimization of the singlet P channel}

In the perturbation theory approach outlined above, we can approximate the occupation of the singlet
P channel in terms of known triple S and D channel wavefunctions.  The associated constraint on
the channel wavefunction can be written as

$$
\bigg [ -{\hbar^2 \over 2 \mu}{d^2 \over dr^2}+{2 \hbar^2 \over 2 \mu r^2} + 9v_C^{os}(r)   - 2 v_{LL}^{os}(r) \bigg ] w(r) 
~=~
M_J
\left ( {1 \over 2 Mc^2} \right )
\left ( {1 \over 2 m_{av} c^2} \right )
(\hbar c) (c \hat{P}_z)
\ \ \ \ \ \ \ \ \ \ \ \ \ \ \ \ \ \ \ \ \ \ \ \ \ \ \ \ \ \ \ \ \
\ \ \ \ \ \ \ \ \ \ \ \ \ \ \ \ \ \ \ \ \ \ \ \ \ \ \ \ \ \ \ \ \
$$
$$
\times
\bigg \lbrace
-2 \sqrt{3}   \left [{d \over dr} v_C^{eT}(r) \right ] u(r) 
+
 12 \sqrt{3}   v_T^{eT}(r) \left [ {d \over dr} u(r) + {u(r) \over r} \right ] 
+
 8 \sqrt{3}    \left [{d \over dr} v_T^{eT}(r)  \right ] u(r) 
$$
$$
+
{2 \over \sqrt{3}}    v_{LS}^{eT}(r) \left [ {d \over dr} u(r) - {u(r) \over r} \right ] 
- 
2 \sqrt{3}   v_{LL}^{eT}(r) \left [ {d \over dr} u(r) - {u(r) \over r} \right ] 
\bigg \rbrace
$$
$$
+
M_J
\left ( {1 \over 2 Mc^2} \right )
\left ( {1 \over 2 m_{av} c^2} \right )
(\hbar c) (c \hat{P}_z)
\bigg \lbrace
- \sqrt{6}   \left [{d \over dr} v_C^{eT}(r) \right ] v(r) 
$$
$$
+ 6 \sqrt{6}   v_T^{eT}(r) \left [ {d \over dr} v(r) + {v(r) \over r} \right ] 
+ 4 \sqrt{6}   \left [{d \over dr} v_T^{eT}(r) \right ] v(r) 
- \sqrt{8 \over 3}    v_{LS}^{eT}(r) \left [ {d \over dr} v(r) + 2{v(r) \over r} \right ] 
$$
\begin{equation}
- \sqrt{6}   \left [ {d \over dr} v_{LS}^{eT}(r) \right ] v(r)  
- 2 \sqrt{6}  v_{LL}^{eT}(r) \left [ {d \over dr} v(r) + 2{v(r) \over r} \right ] 
- \sqrt{6}   \left [ {d \over dr} v_{LL}^{eT}(r) \right ] v(r)  
\bigg \rbrace
\end{equation}

\begin{figure} [t]
\begin{center}
\mbox{\epsfbox{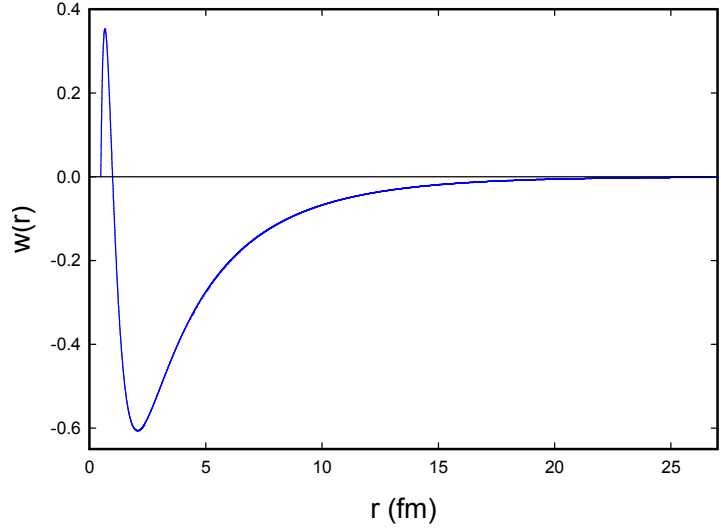}}
\caption{Numerical solution for the normalized singlet P radial wavefunction.}
\label{SingletP}
\end{center}
\end{figure}

\noindent
We have solved this equation numerically assuming that $u(r)$ and $v(r)$ are 
fixed solutions of the Rarita-Schwinger equations, and the resulting normalized
solution for $w(r)$ is shown in Figure \ref{SingletP}.


\epsfxsize = 4.00in
\epsfysize = 3.200in

\subsection{Equivalent two-level model parameters}

From the computation outlined above we can derive an equivalent two-level system
model in the form

\begin{equation}
E
\left ( 
\begin{array} {c}
c_1 \cr
c_2 \cr
\end{array}
\right )
~=~
\left ( 
\begin{array} {cc}
H_{11} & H_{12}  \cr
H_{21} & H_{22}  \cr
\end{array}
\right )
\left ( 
\begin{array} {c}
c_1 \cr
c_2 \cr
\end{array}
\right )
\end{equation}

\noindent
We compute

\begin{equation}
H_{11} ~=~ -2.245~{\rm MeV}
\end{equation}

\begin{equation}
H_{22} ~=~ 125.4~{\rm MeV}
\end{equation}

\begin{equation}
H_{12} ~=~ H_{21} ~=~ 2.98 \times 10^{-3} ~M_J (c \hat{P}_z)
\end{equation}

\noindent
The off-diagonal coupling matrix elements are somewhat smaller than we were hoping for,
and future work will be needed to understand if this coupling is sufficiently large to
account for experimental results.
In addition, we have found that these off-diagonal matrix element depend on the
nuclear spin, which suggests that the system may respond to net spin alignment.


\newpage
\section{Discussion and conclusions}

We recently proposed a new fundamental Hamiltonian for condensed matter
lattice problems that includes coupling to nuclear internal degrees of
freedom.  From our perspective this new coupling seems to be what is
needed to account for the excess heat effect in the Fleischmann-Pons
experiment.  What has been needed in order to evaluate the models that
result is an estimate for the coupling matrix element.

The development of an estimate for this matrix element is challenging
for a variety of reasons.  We have presumed in the derivation of the
fundamental Hamiltonian that it is sufficient to model the nucleons as
elementary Dirac particles.  However, we know that nucleons are composite
particles made up of quarks and gluons, and that it is unlikely that using
a Dirac model as we have done is going to give accurate results.  To
do better we probably need to go back and develop a better fundamental
Hamiltonian based on quarks and electrons.  If it is possible to 
obtain reasonable nucleon models from empirical potentials, then
we may be able to develop a better estimate for the deuteron coupling matrix
element.  Working directly with bound state QCD at this stage does not seem
to be an attractive option.

Once we have decided on the simpler model that adopts an elementary 
Dirac particle model for nucleons, then it is an issue of whether to
use a relativistic or nonrelativistic model, and further it is an
issue of what potential to use.  Since these computations involve a
fair amount of work, it seemed sensible to adopt a nonrelativistic
model since it is simpler, and to work with an older relatively
simple nuclear model.  The Hamada-Johnston potential fits the bill
in this regard, as it is sufficiently simple that we are able to 
complete a calculation in relatively short order.  Perhaps the most
work in this computation was the evaluation of the spin, isospin, and
angular momentum algebra; for this we relied on brute force Mathematica
calculations.

In the end, we have developed a model for the coupling between the
different nuclear spin states of the ground state deuteron and 
lattice-induced coupling to a highly-excited singlet P virtual state.
The energy of this virtual state is about 125 MeV in this model, which
is consistent with our expectations.  The coupling matrix element fell
short of what we had hoped for by about an order of magnitude.  We will
need to clarify in future calculations if this is sufficiently large
to be relevant to experimental results.

The coupling matrix element in this model is proportional to $M_J$, which
is interesting in connection with the reported dependence of excess heat
on the strength of an applied magnetic field.    Since the matrix
element is proportional to $M_J$, there is the potential for a larger coupling
if the deuteron spins can be aligned.  We are interested in pursuing this
possibility in future work.


\end{document}